\documentclass[aps,prl,twocolumn,preprintnumbers,superscriptaddress,letterpaper]{revtex4-1}
\usepackage{graphicx}
\usepackage{epstopdf}
\usepackage{amssymb}
\usepackage{array}
\usepackage{amsmath}
\usepackage{amsfonts}
\usepackage{bm}
\usepackage{color}
\usepackage{tabularx}
\usepackage[hidelinks]{hyperref}
\usepackage[normalem]{ulem}

\definecolor{blueblack}{rgb}{0, 0.2, 0.75}

\definecolor{lightgrey}{rgb}{0.8, 0.8, 0.8}

\begin{document}

	
\title{A lower bound to the thermal diffusivity of insulators}

\author{Kamran Behnia}
\email[Corresponding author: ]{kamran.behnia@espci.fr}
\affiliation{Laboratoire Physique et Etude de Mat\'eriaux (CNRS-Sorbonne Universit\'e), ESPCI Paris, PSL Research University, 75005 Paris, France}
\affiliation {II. Physikalisches Institut, Universit\"{a}t zu K\"{o}ln, 50937 K\"{o}ln, Germany}
\author{Aharon Kapitulnik}
\affiliation{Departments of Physics and Applied Physics, Stanford University, Stanford, CA 94305,USA}
\affiliation{Geballe Laboratory for Advanced Materials, Stanford University, Stanford, CA 94305, USA}

\date{\today}

\begin{abstract}
It has been known for decades that thermal conductivity of insulating crystals becomes proportional to the inverse of temperature when the latter is comparable to, or higher than, the Debye temperature. This behavior has been  understood as resulting from Umklapp scattering among phonons. We put under scrutiny the magnitude of the thermal diffusion constant in this regime and find that it does not fall below a threshold set by  the square of sound velocity times the Planckian time ($\tau_p=\hbar/k_BT$). The conclusion, based on scrutinizing the ratio in  cubic crystals with high thermal resistivity, appears to hold even in glasses where Umklapp events are not conceivable. Explaining this boundary, reminiscent of a recently-noticed limit for charge transport in metals, is a challenge to theory.
\end{abstract}
\maketitle
\section{Introduction}
In crystalline insulators, thermal conductivity, $\kappa$,  peaks at an intermediate temperature. This is a consequence of two competing tendencies. Warming increases the population of heat-carrying  phonons, but also multiplies the rate of flow-degrading collisions.  In the high temperature limit above the peak, phonons are mainly scattered by other phonons. Accurate description of the thermal conductivity of such solids by the Peierls-Boltzmann equation in this intrinsic regime is another remarkable accomplishment of the quantum theory of solids. For example, first-principle theory is nowadays capable of giving a quantitative account of intrinsic thermal conductivity of familiar semiconductors \cite{Lindsay2016}.

In thin paper, we focus our attention to the high-temperature regime. Here, a ubiquitous asymptotic behavior $\kappa \propto T^{-\alpha}$ , with $\alpha \gtrsim 1$ \cite{Berman1976} is visible in the vicinity of the Debye temperature  and above. This behavior is attributed to multi-phonon scattering processes, enabled by anharmonic terms in the Hamiltonian, which allow for entropy to be transferred to the crystal. Since normal (N) processes conserve real momentum exactly, they cannot thermalize the crystal. On the other hand, in umklapp (U) processes, where a non-zero Reciprocal Lattice vector is involved, real momentum, (and thus entropy) is transferred to the center of mass of the crystal as thermal equilibrium is restored.   However, the N collisions, by influencing the momentum distribution among phonons, play an indirect role in setting the magnitude of thermal conductivity \cite{Callaway,Allen2013}. Note that while thermal \textit{resistivity} of insulators tends to become T-linear above the Debye temperature in all insulators, its \textit{magnitude} greatly varies among different solids. In other words, the prefactor of this T-linear behavior is material specific. Cubic diamond  \cite{Onn} conducts heat 500 times more than  cubic lead telluride \cite{delalonde}.

In the last century much attention was devoted to the link between the mean-free-path for phonon-phonon scattering and specific material-related properties such as Gr\"uneisen parameter,  Debye temperature, lattice spacing and atomic mass \cite{Berman1976,Leibfried,Dugdale,Julian,Slack1973,Slack1979}. The driving idea is sketched  by stating that departure from harmonicity sets both the Gr\"uneisen parameter and the rate of phonon-phonon collisions\cite{Dugdale,Berman1976}. According to the data gathered by Slack \cite{Slack1973,Slack1979}, this approach is successful in explaining the order of magnitude of the thermal conductivity of non-metallic crystals,  but only when the phonon mean-free-path exceds the lattice constant or the phonon wavelength \cite{Slack1979}.  More recently, attention has turned to the role of other features, such as resonant bonding  \cite{Matsunaga2011,Lee2014} and the phase space available for three-phonon scattering \cite{Lindsay2008,Fu2018}. 
In principle, these can be quantified from \textit{ab initio} phonon spectrum  \cite{Lindsay2008,Fu2018}. 

Electrical resistivity in common metals becomes $T$-linear in the vicinity of Debye-temperature and above. This behavior is what is expected according to the Bloch-Gr\"uneisen picture of electron-phonon scattering \cite{Ashcroft}. Caused by strong electron-electron interactions, T-linear resistivity is also present in the so-called ``bad'' metals \cite{Emery}, such as high-T$_c$ cuprates \cite{Zaanen}. T-linear resistivity is thought to be a signature of quantum criticality \cite{Sachdev2011} and  has been indeed observed in numerous strongly-correlated electron systems pushed near a quantum critical point. A careful examination by Bruin \textit{et al.} \cite{Bruin2013} found that in all these cases, the average scattering time is close to the Planckian dissipative timescale: $\tau_p=\frac{\hbar}{k_{B}T}$  \cite{Zaanen,Sachdev2011}. Following this striking observation, Hartnoll proposed that there may be a lower bound boundary to diffusive constant of electrons propagating in a metal, D$_{qp}$ \cite{Hartnoll2015} :

\begin{equation}\label{eqn1}
D_{qp}\gtrsim \frac{\hbar v_{F}^{2}}{k_{B}T}=v_F^2\tau_p
\end{equation}
where $v_F$ is the Fermi velocity.

In this paper, we focus on the magnitude of the thermal diffusivity in insulators, $D_{th}$ and find that the  planckian disspiaptive time-scale  is relavant to phonon thermal transport.

\section{Thermal diffusivity}
Thermal diffusivity, $D_{th}$ is defined by the heat equation:
\begin{equation}\label{Heat}
\frac{\partial T}{\partial t}= D_{th} \frac{\partial^2T}{\partial x^2}
\end{equation}

Its dimensional unit is square of length divided by time. Thermal conductivity, $\kappa$, on the other hand, is defined by the  Fourier equation,  the thermal equivalent of the Ohm's law, which relates the heat current density to the thermal gradient:

\begin{equation}\label{Fourrier}
J_q= -\kappa \frac{\partial T}{\partial x}
\end{equation}

Conservation of energy implies:
\begin{equation}\label{Conserv}
\nabla \cdot J_q= - \frac{\partial E}{\partial t}
\end{equation}

Combining the last two equations with the definition of specific heat per volume, $C=\frac{\partial E}{\partial T}$, leads to a fundamental relation between $\kappa$, $C$ and $D_{th}$.
\begin{equation}\label{kappaD}
 D_{th}= \frac{\kappa}{C}
\end{equation}

In general, thermal conductivity is a sum involving many phonon modes (indexed $\lambda$); a product of their specific heat, their velocity and the distance they travel between two collisions:

\begin{equation}\label{kappaD}
\kappa= \sum_{\lambda} C_{\lambda}v_{\lambda}\ell_{\lambda}
\end{equation}

If averaging is not skewed,  $D_{th}$ is simply the product of the mean velocity, $v$, and the mean free path, $\ell$: $D_{th}=\tfrac{1}{d}v\ell$ ($d$ is the relevant dimensionality).

Below, we will put under scrutiny the available data on a variety of insulators, we find that the thermal diffusivity never falls below a threshold value, respecting an inequality similar to inequality \ref{eqn1}, with Fermi velocity replaced by the sound velocity, $v_s$. We will specifically focus on the velocity of the  longitudinal acoustic (LA) mode, which involves excursions of atoms along the direction of heat propagation and thus is most efficient in transporting entropy.
\begin{equation}
\label{eq2}
D_{th} \gtrsim  \frac{\hbar v_s^2}{k_BT}=v_s^2\tau_p
\end{equation}

Interestingly, this represents the boundaries of the adiabatic approximation in the semiclassical picture \cite{Ziman}. 

More specifically, the high-temperature thermal diffusivity, $\bar{D}_{th}$  is governed by the sound velocity $v_s$, and the Planckian relaxation time, $\tau_p$, such that
\begin{equation}
\label{eq3}
\bar{D}_{th} =sv_s^2\tau_p
\end{equation}
where $s >1$ is a constant specific for different families of materials. To demonstrate the broad validity of this bound,  we will show here that this observation (hinted previously in \cite{Zhang2017,Martelli,Zhang2018}) holds in two distinctly different families of low-conductivity cubic crystals as well as in an amorphous solid. We will also argue that the relevance of this inequality brings additional insight to specific peculiarities in a given family. We will end with a brief discussion on a possible link between this observation and the uncertainty principle in the context of random walks.

\section{Cubic crystals}
\begin{figure*}
\centering
\includegraphics[width=1.3\columnwidth]{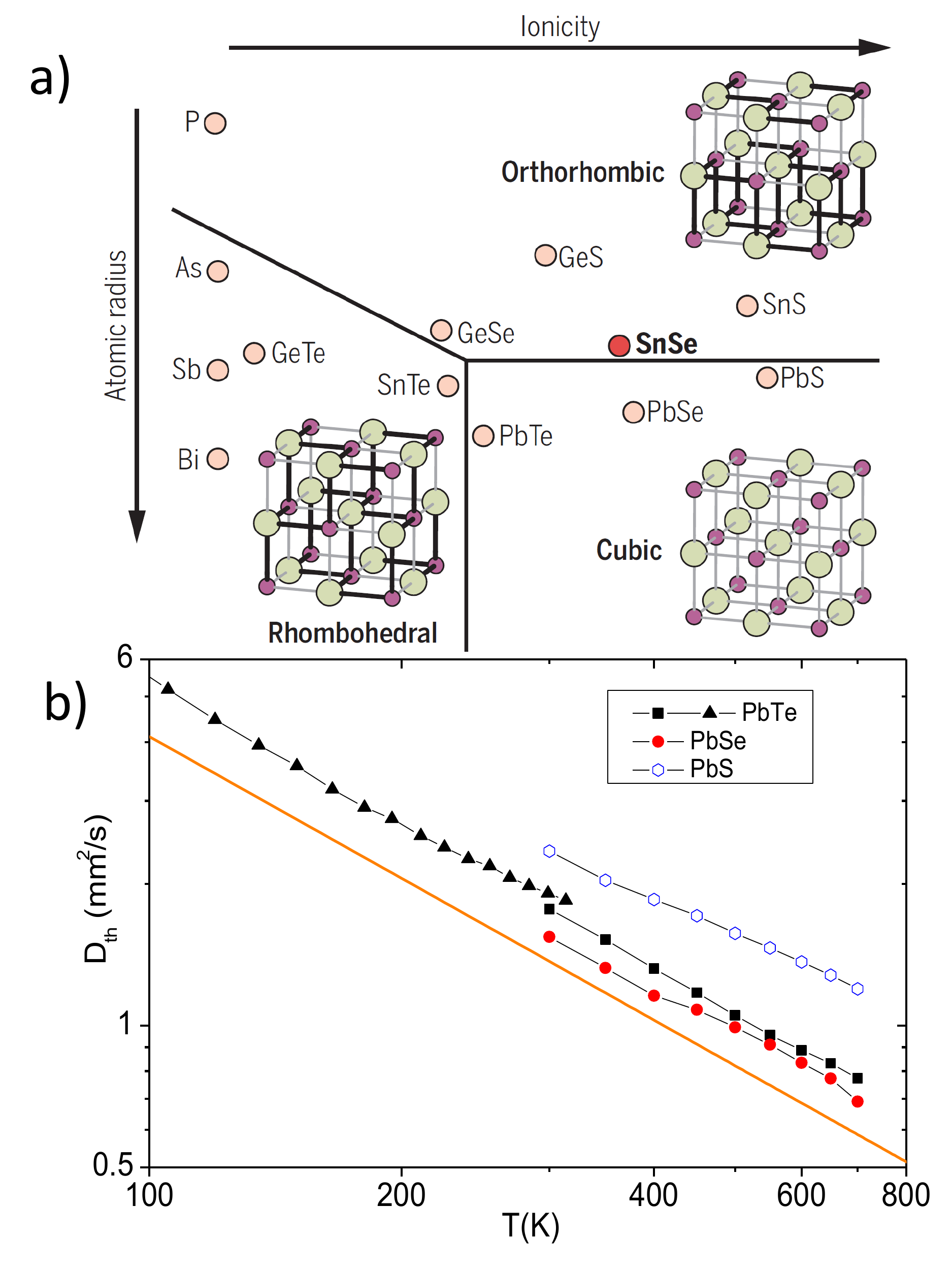}
\caption{a) Structural phase diagram of the column-V elements and IV-VI semiconductors \cite{Littlewood,Behnia2016}.  Many of these binary salts, known to have a very low thermal conductivity, have been explored as potentially interesting thermoelectric materials. The  cubic members of this family  are PbS, PbSe and PbTe. b) Temperature dependence of thermal diffusivity in these three solids extracted from the reported data \cite{Parkinson,Pashinkin,Morelli,elsharkawy} on thermal conductivity and specific heat. Two distinct symbols were used for the independent sets of data for PbTe. The orange solid line represents $ s v_s^2 \tau_p $, with v$_s$=3.0 km/s \cite {Jacobsen} and s=6. }
\label{cubic}
\end{figure*}

In anisotropic materials, particularly in a direction along van-der-Waals bonding, display low heat and charge transport conductivity.  Where itinerant electrons are present, the electrical conductivity usually exhibits a much stronger anisotropy than either the thermal conductivity, or the relevant longitudinal sound velocity. However, the anisotropy of these last two quantities which involve phonons are far from being negligible. For example, for graphite while  $\sigma_{ab}/\sigma_c > 10^4$ \cite{Krishnan1939}, $\kappa_{ab}/\kappa_{c} >50$ \cite{Slack1962} and $v_s^{ab}/v_s^c > 5.3$ \cite{Bosak2007}, where for the anisotropic strongly-correlated cuprate Bi$_2$Sr$_2$CaCu$_2$O$_8$, while $\sigma_{ab}/\sigma_c > 10^5$ \cite{Martin1988}, $\kappa_{ab}/\kappa_{c} >7.1$ \cite{Wu1993}, and  $v_s^{ab}/v_s^c > 1.6$ \cite{Saunders1994}. 

We will focus here on cubic solids where thermal conductivity and diffusivity does not depend  on the orientation of heat propagation. In particular, we will consider the magnitude and temperature dependence of two families of cubic solids known for their exceptionally low thermal conductivity.

\textbf{Cubic IV-VI semiconductors} - These  binary salts contain elements of column IV and column VI of the periodic table. The IV-VI family and column V elements crystalize in one of the three varieties derived from the cubic rocksalt (See Fig.~\ref{cubic}a).  Tin selenide and black phosphorous are orthorhombic, but bismuth and tin telluride are rhombohedral. It is known that compared to the III-V semiconductors (such as GaAs) or Column IV elements (such a diamond), these solids conduct heat much less \cite{Lee2014}. This unusually low thermal conductivity have made members of this family attractive thermoelectric materials \cite{Heremans,Lidong}.

Fig.~\ref{cubic} shows the high-temperature thermal diffusivity of three cubic members of this family, namely PbTe, PbSe and PbS using a variety published data on specific heat \cite{Parkinson,Pashinkin,elsharkawy} and thermal conductivity \cite{Morelli,elsharkawy}. One can see that between 100 K and 600 K, the amplitude of $D_{th}$ is  not very different among these three solids and its temperature dependence remains close to $T^{-1}$.  In fact, using a value of $s=6$ in Eqn.~\ref{eq3} and the longitudinal sound velocity of PbTe \cite {Jacobsen}, we see that all the data of this class of materials fall above this line (see orange line in Fig.~\ref{cubic}b).

Note that the Debye temperature in these solids is low and even the energy of the optical phonons is below 15 meV \cite{An,Ribeiro}. Thus, in the temperature range of interest, all phonons are thermally excited and the phonon gas may be considered a classical gas of particles obeying the Maxwell-Boltzmann statistics \cite {footnote1}.  Since cubic IV-VI semiconductors are considered as the worst heat conductors among crystals,  one may assume that in other insulating heat conductors  the relaxation time will be even larger compared to the Planckian time (i.e. $s > 6$). However, as we will see below, it is the Diffusivity, that is, the combination of the velocity and relaxation time that determine the ability of a phonon system to thermalize and higher thermal conductivity may still correspond to a closer proximity of the relaxation time to the Planckian limit.

\begin{figure*}
\centering
\includegraphics[width=1.0\columnwidth]{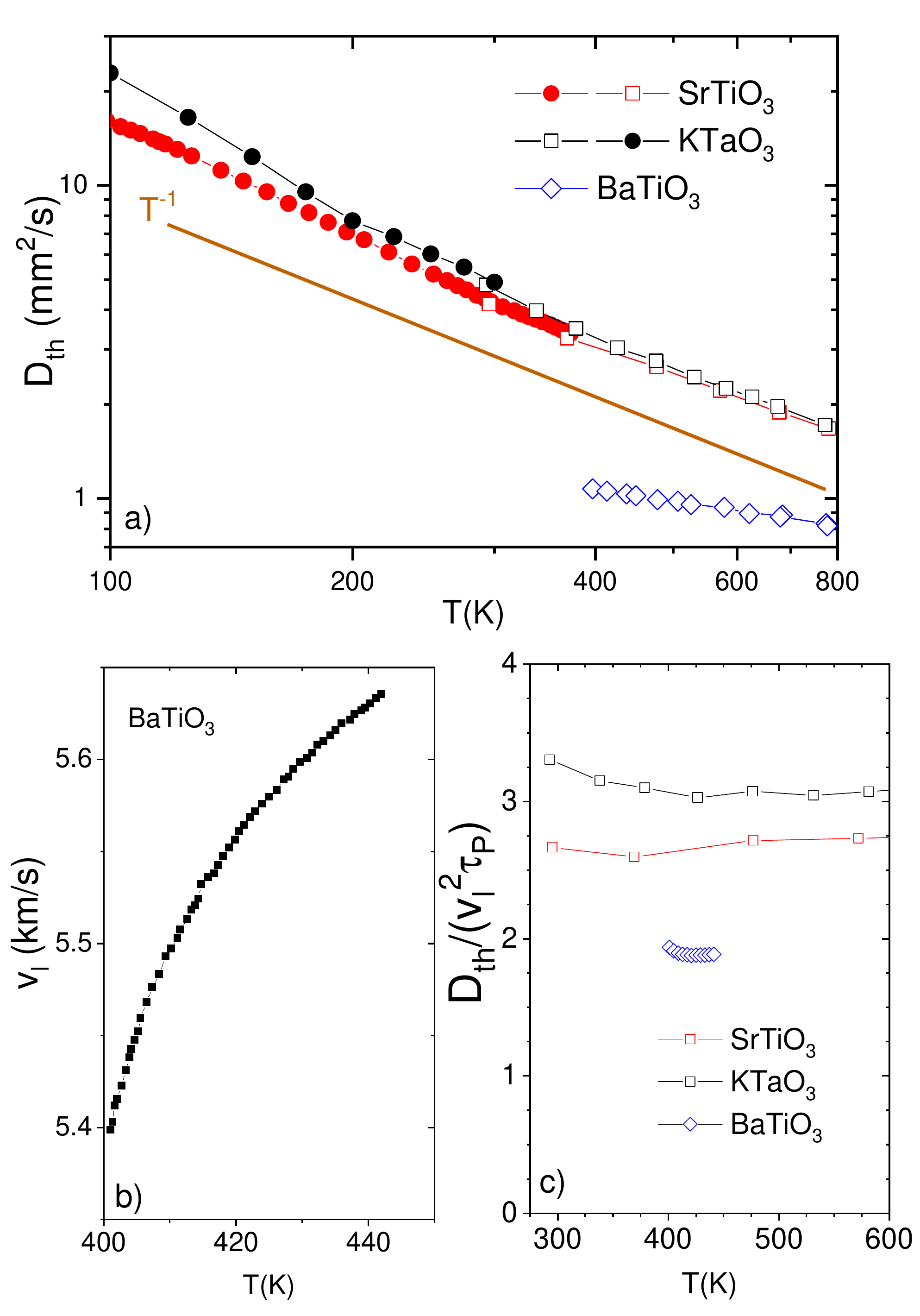}
\caption{a) Temperature dependence of thermal diffusivity in three cubic perovskites. High-temperature data for SrTiO$_3$  (open red squares),  KTiO$_3$ (open black squares) and BaTiO$_3$ (open blue diamonds) were reported in ref. \cite{Hofmeister}. Low-temperature data for  SrTiO$_3$  (solid red circles) are from data reported in ref.\cite{Martelli}. Low-temperature data for KTiO$_3$ (solid black circles) were extracted by taking the thermal conductivity reported in ref.\cite{Tachibana} and specific heat reported in ref.\cite{White}. Solid orange line represents T$^{-1}$ temperature dependence. Note that BaTiO$_3$ looses its cubic symmetry below 400 K. b) Temperature dependence of the sound velocity in BaTiO$_3$ as reported in ref.\cite{Kashida}. This unusually large temperature dependence may  explain why the temperature dependence of thermal diffusivity is slower than T$^{-1}$, as seen in the next panel. c) Thermal diffusivity  divided by the square of sound velocity and $\tau_P$ in the three perovskites. The longitudinal sound velocity is 7.5 km/s in KTaO$_3$ \cite{Barrett} and 7.9 km/s in SrTiO$_3$ \cite{Okai}. In all three cases, this dimensionless ratio is flat and larger than unity.}
\label{sto}
\end{figure*}
\bigskip

\textbf{Cubic perovskites} -  This family of solids is abundant among oxides having the generic chemical formula  ABO$_3$. Because of their  larger thermal conductivity compared to the IV-VI family, their potential for thermoelectric application \cite{Okuda} is limited. For example, room-temperature thermal conductivity is 13 W/K$\cdot$m in SrTiO$_3$ \cite{Martelli} and 16 W/K$\cdot$m in KTaO$_3$ \cite{Tachibana} compared to 2 W/K$\cdot$m in PbTe \cite{Morelli}. Thus, one expects the thermal diffusivity of the latter to be larger by a similar factor. As one can see in Fig.~\ref{sto}, this is basically true. Thermal diffusivity of these materials above room temperature has been extensively studied by Hofmeister \cite{Hofmeister}. Her results are in very good agreement with the specific heat and thermal conductivity results at room temperature and below reported by other authors \cite{Martelli,Tachibana}.  The room temperature magnitude is about 5 mm$^2$/s compared to 2 mm$^2$/s in PbTe. However, since the sound velocity in the two perovskites is significantly larger, their relaxation time is closer to the Planckian time.

Another feature that is observed in Fig.~\ref{sto} is that (at least above 200 K) D$_{th}$ follows a  $T^{-1}$ temperature dependence in both SrTiO$_3$ and KTaO$_3$ and  many other oxides \cite{Hofmeister}). Note that in these materials, in contrast to the previous case, the temperature range of interest remains below the energy of the largest optical phonon, which is as high as 100 meV \cite{Aschauer}. Nevertheless, the temperature dependence of the diffusivity is still close to T$^{-1}$. This feature may suggest that the limit of Planckian time is a generic feature of the dynamics rather than the thermodynamics of the phonon system.

It is instructive to examine the specific case of BaTiO$_3$ in more detail. This archetypal ferroelectric \cite{Vija} is cubic above 400 K and passes through three successive structural transitions upon cooling. Its D$_{th}$ in the cubic phase increases with decreasing temperature but does not follow a T$^{-1}$ behavior, as one can see in Fig.~\ref{sto}a. At the same time, the sound velocity in BaTiO$_3$ \cite{Kashida} shows an unusually large temperature dependence near the 400 K structural instability (Fig.~\ref{sto}b ). Taking into account this temperature dependence, one finds that $D_{th}/v_s^2$ is proportional to T$^{-1}$. In other words, as one can see in Fig.~\ref{sto}c, according to the available data, D$_{th}T/v_s^2$ is flat in the three perovskites. This supports the idea that phonon scattering  is intimately linked to $\tau_p$ times a factor of the order of unity. Further measurements over a broader temperature window are desired to confirm this picture.

\section{glasses}
\begin{figure}
\includegraphics[width=1.0\columnwidth]{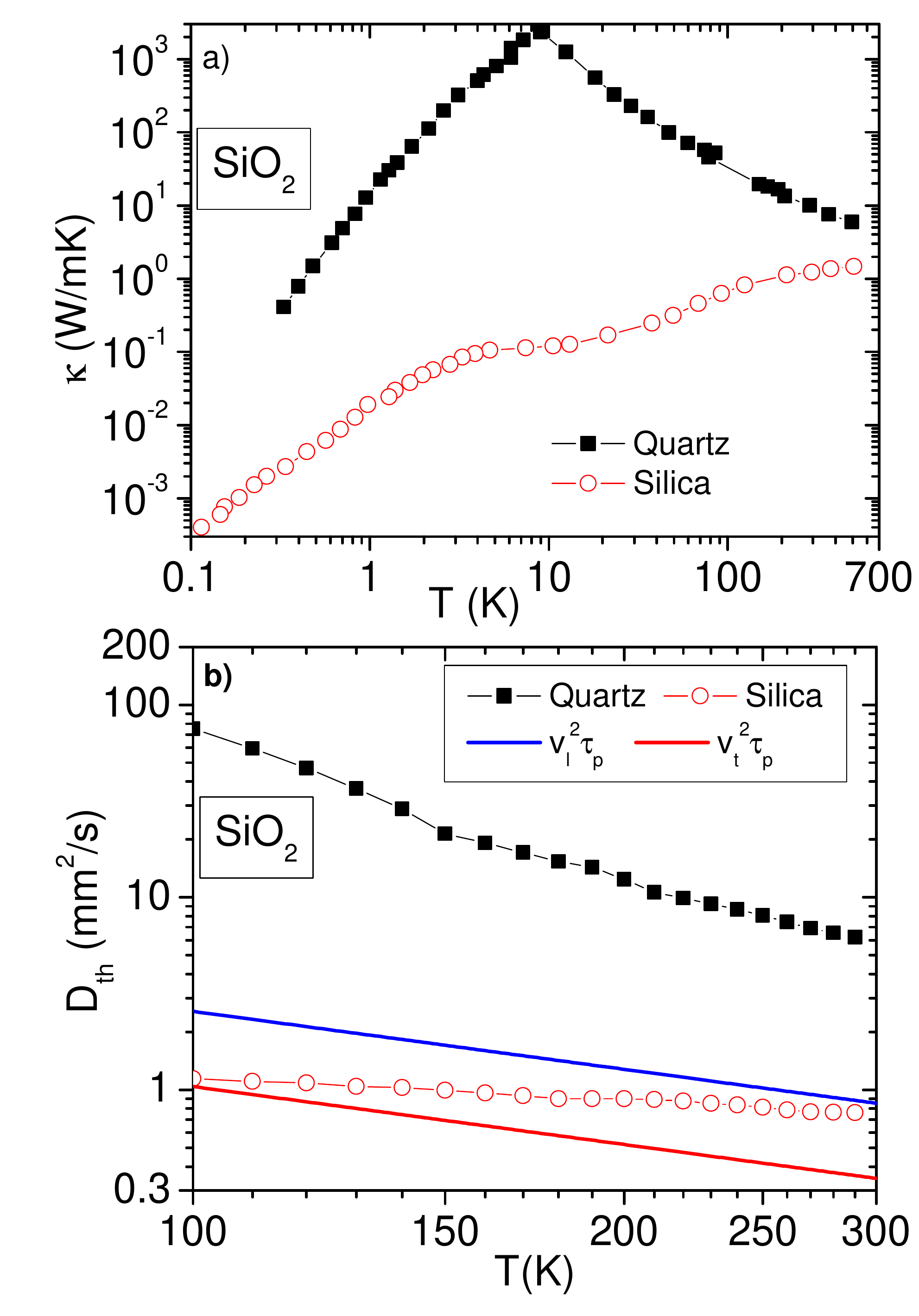}
\caption{a)Temperature dependence of thermal conductivity  in crystalline SiO$_2$ (quartz)  and amorphous SiO$_2$ (silica) \cite{Vandersande,Zeller}.  b)Temperature dependence of thermal diffusivity of the two systems extracted from thermal conductivity and specific heat data reported by Zeller and Pohl \cite{Zeller}. No report on specific data above room temperature was found and this is the reason for the restricted temperature range. Blue and red solid lines represent $v_l^2\tau_p$ and $v_t^2\tau_p$, using the reported values of longitudinal ($v_l$=5.8km/s) and transverse ($v_t$=3.7km/s) sound velocities of silica \cite{Pohl}.}
\label{glass}
\end{figure}
 In crystalline materials, lattice vibrations and their phonon ``normal modes'' are well-defined. It is understood that anharmonicity lead to a finite relaxation time.  By contrast, normal modes in glasses are no longer simple plane-wave phonons. While a well defined dispersion relation exists, and both long wavelength longitudinal and transverse sound waves can be measured in glassy solids, the extracted phonon mean-free-path  becomes of order of the (average) lattice parameter \cite{Kittel}. Note that (like cubic crystals) a perfect glaasy solid is expected to be isotropic.

There have been several attempts to provide an adequate description of heat propagation in glasses. For example, Allen and Feldman \cite{Allen93}proposed a theoretical framework of heat transport for such solids in which heat is mostly carried by vibrational modes that are neither localized nor propagating. However, to the best of our knowledge, no bound has been set for the magnitude of thermal diffusivity.

Before examining the relevance of inequality~\ref{eq2} to the experimental data, let us note the qualitative difference between the temperature dependence of thermal conductivity in glasses and in crystals \cite{Vandersande}. This can be seen in Fig.~\ref{glass}a, where we reproduce the published  thermal conductivity data in SiO$_2$. One can see that the thermal conductivity of the amorphous solid, instead of displaying a peak in its temperature dependence, is monotonously increasing with warming.

Fig.~\ref{glass}b shows the temperature dependence of thermal diffusivity extracted from thermal conductivity and specific heat data. Since there is no report on the evolution of the latter above room temperature, the temperature range is restricted. The figure inspires three observations. First of all, D$_{th}$ of silica (the glassy SiO$_2$) is mildly \textit{decreasing} with warming. Second, this decrease is slower than T$^{-1}$. Our third observation is that in this temperature range, the amplitude of diffusivity is within the boundaries set by the product of $\tau_p$ and the square of longitudinal and transverse velocities reported in the review paper by Pohl and co-workers \cite{Pohl}. We note that there is no report on any study of the temperature dependence of the sound velocity in this amorphous solid.

Thus, while the amplitude of thermal diffusivity in glasses is tantalizingly close to the limit set by inequality~\ref{eq2}, at this point the available data does not allow to draw a definite conclusion regarding its strict validity.

We also note that at high enough temperature thermal conductivity of glasses saturates to a constant value\cite{Pohl}. The Dulong-Petit law implies a similar saturation for specific heat. Therefore, ultimately, thermal diffusivity will tend towards an asymptotic magnitude, D$_{th}^\infty$ of the order of v$_sa$ (where $a$ is a length of the order of ineteratomic distance). Extrapolating the data (see Fig. 3b)  indicates that D$_{th}^\infty$ in this regime will still exceed  $v_l^2\tau_p$ and $v_t^2\tau_p$.  

\section{Concluding remarks}
We saw that thermal diffusivity of crystals in the high-temperature regime has a simple temperature dependence, which can be explained if one assumes that heat carriers are diffusive quasi-particles with the velocity of sound and  a scattering time equal to the Planckian time multiplied by a numerical factor. The numerical factor is different for different types of materials and for some it approaches unity, but does not fall below it. This suggests the existence of a universal bound to thermal diffusivity.

Such an approach based on thermal diffusivity brings interesting insights.  In the case of BaTiO$_3$ above  Curie temperature, it links the unusually slow temperature dependence of thermal conductivity  to the unusually large temperature dependence of sound velocity. In a comparison of thermal transport in two distinct family of solids (IV-VI and ABO$_3$), it identifies the low magnitude of sound velocity as the origin of low thermal conductivity.

As an example of additional insight brought by this approach, consider silver halides, AgCl and AgBr proposed by a recent theoretical study \cite{Yang19} to be record-breaking thermoelectrc materials. Theory expects them to display a thermal conductivity as low as 0.2 W/Km at 300 K and 0.09 W/Km at 600 K \cite{Yang19}. This would mean a room-temperature heat conductivity in these crystals four times lower than in amorphous silica. In our picture, however, such a low thermal conductivity is allowed because the sound velocity of silver halides is low \cite{Takesawa} and their lattice parameter large. Therefore, even if future experiments confirm the remarkably low thermal conductivity predicted by theory, they would not violate the bound on thermal diffusivity discussed here \cite{footnote2}.

Heat transport of crystalline solids in this temperature window has been traditionally ascribed to Umklapp scattering of phonons off each other. However even amorphous solids (which do not host Umklapp events) appear to respect the thermal diffusivity bound in a loosely fashion. These observations suggest a more fundamental reason for this observed bound.

The presence of the Planck constant implies that any explanation should be  quantum mechanical, even at these high temperatures. According to the Heisenberg uncertainty principle, particles with a thermal energy of $k_B T$ cannot be well defined within a time-scale shorter than $\delta t= \hbar/K_B T$. Moreover, if they have a velocity of $v$ their thermal momentum would be $P_{th}=\frac{k_B T}{v}$. Their position  cannot be defined with an accuracy better than $\delta x=\hbar v/k_{B}T$. 

On the other hand, the equation describing the flow of heat in a dense medium (Eq.\ref{Heat}) is a classical equation. Yet, it contains a tension between its apparent continuity and the fundamental discreteness of the random walk process generating it \cite{Lawler}. This tension between discreteness and continuity survives in more sophisticated treatments of the passage between random walk and diffusive propagation \cite{Bouchaud}. Now, thermal diffusion generated by random  walk of particles with a typical velocity of $v$, implies a lower limit to accuracy. This is true both for position (undefined better than  $\delta x=\frac{D_{th}}{v}$) and for time (within $\delta t=\frac{D_{th}}{v^2}$). If one assumes that these time and length scales match those imposed by the quantum mechanics, then one deduces inequality \ref{eq2}. 
A possible way to connect the continuous heat equation to the discrete random walk is to invoke the minimum coarse-grained phase space available to any system, which is $\delta x \delta P=\hbar$. While the thermodynamics may be considered in the classical limit, the dynamics, that is, the exploration of states in phase space while achieving thermalization may still sense the underlying quantum-mechanical ``grid''. 

We hope that our observation stimulates further theoretical \cite{Davison,Blake,Patel,ChenLin} and experimental investigations.

\section*{ Acknowledgments}

We thank Sean Hartnoll, Steve Kivelson and Rapha\"el Voituriez for  discussions. This work was initiated through a ``QuantEmX'' Exchange Awards at Stanford University (KB) and at ESPCI (AK). Work at Stanford University was supported by the Gordon and Betty Moore Foundation through Emergent Phenomena in Quantum Systems (EPiQS) Initiative Grant GBMF4529, and by the U. S. Department of Energy (DOE) Office of Basic Energy Science, Division of Materials Science and Engineering at Stanford under contract No. DE-AC02-76SF00515. Work at University of Maryland was supported by NSF grant DMR-1708334. Work in ESPCI was supported by by the Agence Nationale de la Recherche (ANR-18-CE92-0020-01).




\begin{references}
\bibitem{footnote1}
Nevertheless, the  Planck constant remains relevant. It quantifies the unit cell of the phase space and shows up in the Sackur-Tetrod entropy.
\bibitem{footnote2}
Assuming a heat capacity matching the Dulong-Petit value, a 600 K thermal conductivity of 0.09 W/k.m would correspond to a diffusivity as low as 0.05 mm$^2$/s. But with a sound velocity of 1600 m/s \cite{Takesawa}, the scattering time would be 20 fs, which exceeds the Planckian time at 600 K (12 fs).
\end{references}
\end{document}